# (Non-)retracted academic papers in OpenAlex

## Christian Hauschke[1*], Serhii Nazarovets[2]

[1] TIB – Leibniz Information Centre for Science and Technology, Hannover, Germany; christian.hauschke@tib.eu

[2] Borys Grinchenko Kyiv Metropolitan University, 18/2 Bulvarno-Kudriavska Str., 04053 Kyiv, Ukraine; serhii.nazarovets@gmail.com

* Corresponding author

**Abstract**

The proliferation of scholarly publications underscores the necessity for reliable tools to navigate scientific literature. OpenAlex, an emerging platform amalgamating data from diverse academic sources, holds promise in meeting these evolving demands. Nonetheless, our investigation uncovered a flaw in OpenAlex's portrayal of publication status, particularly concerning retractions. Despite accurate metadata sourced from Crossref database, OpenAlex consolidated this information into a single boolean field, "is_retracted," leading to misclassifications of papers. This challenge not only impacts OpenAlex users but also extends to users of other academic resources integrating the OpenAlex API. The issue affects data provided by OpenAlex in the period between 22 Dec 2023 and 19 Mar 2024. Anyone using data from this period should urgently check it and replace it if necessary.

*Introduction*

The exponential growth of scholarly publications highlights the increasing need for tools that facilitate rapid access to current and authentic scientific knowledge. Such tools not only aid researchers in staying up-to-date of the latest advancements but also play a pivotal role in conducting bibliometric analyses, thereby enabling the evaluation of the evolution of scientific literature within different domains. These analyses serve as crucial metrics for assessing the productivity and impact of authors, institutions, and journals. Among the emerging online resources in this domain stands OpenAlex[1], a noteworthy platform known for its openness and data integration capabilities. OpenAlex consolidates and standardizes data from diverse academic sources, with notable emphasis on the Microsoft Academic Graph, which ceased operation in December 2021 (Scheidsteger & Haunschild, 2023), and the extensive corpus maintained by Crossref[2], the largest DOI registration agency (Singh Chawla, 2022).

The current scholarly communication landscape is witnessing a significant shift towards open science (Liu & Liu, 2023). In this evolving paradigm, OpenAlex by OurResearch emerges as a solution that is better aligned with the current requisites of the academic community when compared to the closed, subscription-based citation databases such as Web of Science and Scopus. OpenAlex provides a significantly broader coverage of academic literature, as noted by Priem et al. (2022) and Scheidsteger & Haunschild (2022), thereby addressing the growing demand for comprehensive and accessible sources of research information. Moreover, the OpenAlex API presents a compelling advantage with its unrestricted access to metadata retrieval, rendering it an invaluable resource for conducting large-scale bibliometric analyses (Harder, 2024; Velez-Estevez, 2023). Furthermore, the provision of database snapshots empowers users with the capability to obtain full copies of the OpenAlex database for deployment on their own servers, thereby enhancing accessibility and facilitating further research endeavours.

---

[1] https://openalex.org/
[2] https://help.openalex.org/how-it-works/entities-overview

Since its launch in January 2022, OpenAlex has swiftly garnered substantial interest among academic stakeholders. A notable illustration of this phenomenon is exemplified by Sorbonne University, which, in alignment with its overarching policy of fostering openness, opted not to renew its subscription to the Clarivate bibliometric tools. Instead, the university redirected its focus towards exploring open tools alternatives, with OpenAlex emerging as a prominent candidate[3]. Similarly, the Center for Science and Technology Studies (CWTS) at Leiden University has integrated OpenAlex as a cornerstone data source for its novel CWTS Leiden Ranking Open Edition initiative. This pioneering endeavour aims to equip stakeholders with "fully transparent information about the scientific performance of over 1400 major universities worldwide".[4]

The rapid evolution of computer technology has enabled us to swiftly combine bibliographic data from diverse sources and automate its processing and analysis. However, while such advancements offer immense potential, they often entail challenges concerning the accuracy, comprehensiveness, and standardization of data obtained from disparate sources. As a new data source, OpenAlex faces precisely these challenges. A recent comprehensive investigation conducted by Zhang et al. (2024) delves into the issue of absent affiliations in the metadata of journal articles within the OpenAlex platform. Analysis by Jahn et al. (2023) found that the is_oa filter in OpenAlex, which indicates the availability of open full texts, did not always match the open access status information of the paper. In this paper, we present our own observations regarding the incorrect representation of retractions within OpenAlex metadata and propose potential remedies to mitigate this issue.

The growing volume of scientific output is accompanied by a corresponding increase in various forms of academic misconduct, including paper mills, questionable journals, plagiarism, and the fabrication or falsification of research findings (Else & Van Noorden, 2021; Freiermuth, 2023; Joelving, 2024; Kendall & Teixeira da Silva, 2024). This concerning trend places heightened demands on journal editors and reviewers, whose workload is experiencing a corresponding escalation (Piniewski et al., 2024). As a result, errors or misconduct may not always be promptly identified. Consequently, there has been a surge in retractions worldwide — a process in which journal editors formally notify readers of publications containing significant flaws or erroneous data, thereby announcing that the reliability of their findings and conclusions is questionable (COPE Council, 2019; Mallapaty, 2024; Rivera & Teixeira da Silva, 2021).

The process of retracting a publication involves a meticulous and exhaustive investigation by the journal's editors, culminating in a formal decision to retract the article. Information about retractions is typically published separately within the journal, where editors explain the rationale behind the decision as well as the date of retraction. For detailed information on retractions of scientific articles, researchers can leverage the Retraction Watch database[5]. Notably, in September 2023, Crossref, the pre-eminent DOI registration agency, acquired the Retraction Watch database[6]. This acquisition enhances the database's utility and accessibility as an important resource for scholarly inquiry.

---

[3] https://www.sorbonne-universite.fr/en/news/sorbonne-university-unsubscribes-web-science
[4] https://open.leidenranking.com/
[5] http://retractiondatabase.org/

Retracted papers are accessible to readers on the journal's website, but they must contain a clear note indicating their retracted status. This serves as a cautionary measure to alert users to potential issues associated with the respective paper. However, ensuring consistent marking of retractions across all reference databases where the publication is indexed remains a challenge. Although it is important that retractions are accurately marked, there are inconsistencies in the way that many databases approach this task (Hesselmann et al., 2017; Vuong, 2020). Therefore, we conducted an investigation to assess how information pertaining to retractions is presented in the metadata of publications within the OpenAlex database.

*Method*

In the initial phase of our study on March 6, 2024, we utilized the OpenAlex API to retrieve 47,720 retraction records[7]. Subsequently, we downloaded these records as a CSV file for further analysis. Upon scrutinizing the obtained results, it became apparent that not all entries designated as retractions were accurate. Closer examination of the OpenAlex metadata revealed that the "is_retracted" field serves as the determinant of a publication's status, with values restricted to either true or false.

As previously mentioned, OpenAlex primarily sources its data from Crossref database[8]. Following Crossref's acquisition of the Retraction Watch database, information from this database was integrated into the Crossref Labs API, accessible through the "update-nature" field[9]. We enriched of 47,018 entries (excluding 704 records lacking DOIs) OpenAlex records with the "update-nature" from Crossref using a Python script. Due to the experimental character of the Labs API it was not possible to get a complete dataset. This resulted in a subset of 20486 records.

*Results and discussion*

The results of our analysis of a subset of the "update-nature" field in Crossref metadata are depicted in Figure 1. It is evident from the figure that this field encompasses a range of classifications beyond retractions, including Corrections, Expressions of Concern, and Crossmark Retractions. Our findings indicate that Crossref presents the publication status granularly in the metadata (as illustrated in Figure 2), but OpenAlex employs an approach that consolidates this information in a single boolean field labelled "is_retracted" (Figure 3). Consequently, the mere presence of any information about an update causes OpenAlex to categorise the publication as retracted.

This representation of publication status in OpenAlex is a significant concern, particularly given the platform's increasing importance. For instance, in our examination of retractions within OpenAlex, we observed that among the most cited papers with a retraction status is a seminal work by Corman et al. (2020), which presented the establishment of an RT-PCR test for the detection of the 2019-nCoV virus, which caused the COVID-19 pandemic. Although this paper underwent minor corrections, it was never retracted. Mislabelling such influential publications as retractions not only has the potential to misinform healthcare professionals and jeopardize patient care but also risks undermining public trust in the quality of scientific research as a whole.

---

[6] https://retractionwatch.com/2023/09/12/the-retraction-watch-database-becomes-completely-open-and-rw-becomes-far-more-sustainable/
[7] https://explore.openalex.org/works?page=1&filter=is_retracted%3Atrue&sort=cited_by_count%3Adesc
[8] https://help.openalex.org/faq
[9] https://doi.org/10.13003/c23rw1d9

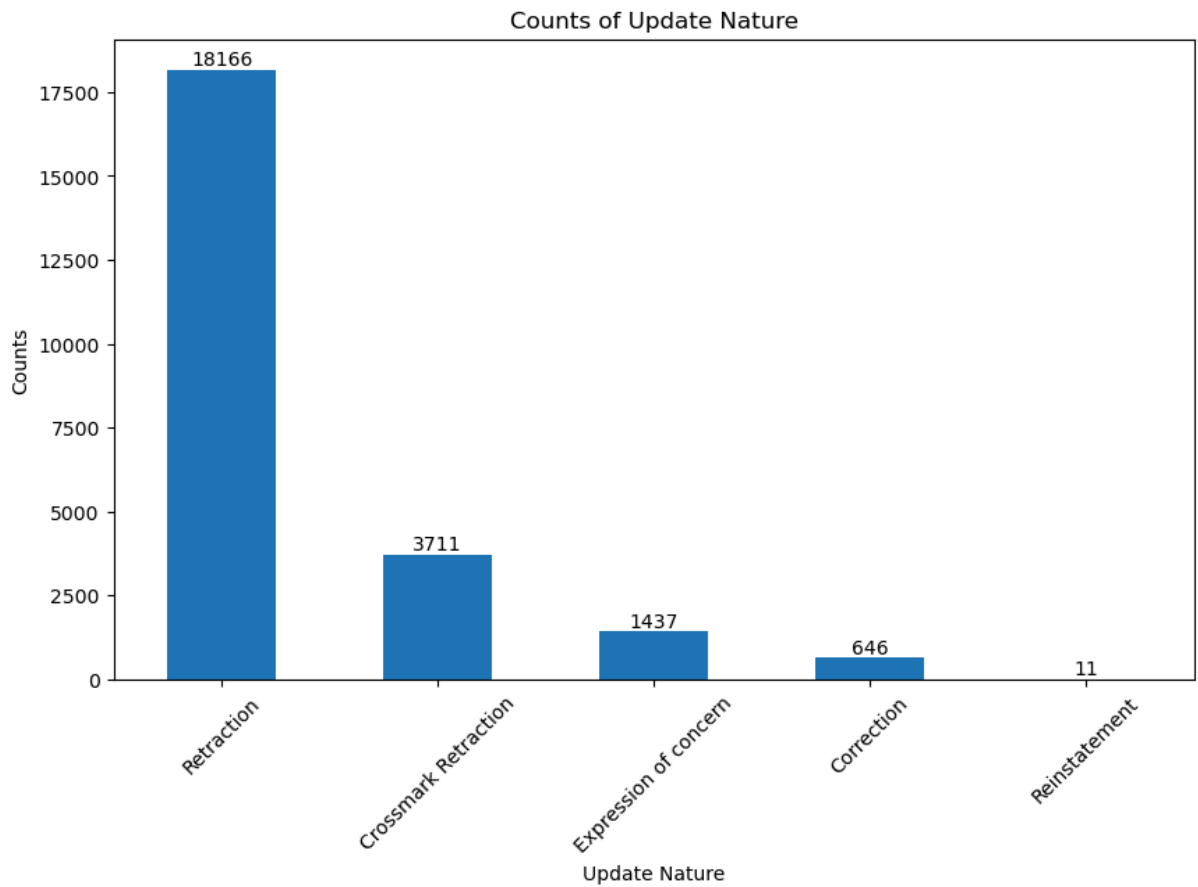

Fig. 1 Results of analysing the content of the "update-nature" field in selected Crossref metadata

```
status:                "ok"
message-type:          "work"
message-version:       "1.0.0"
▼ message:
  ▼ cr-labs-updates:
    ▼ 0:
      ▼ about:
        ▼ source:      "This work has an update record associated with it, asserted by Retraction Watch."
          source_url:  "https://retractionwatch.com"
        ▼ stability:   "The keys used in this API block are unstable and subject to change at any future time."
          asserted-by: "https://ror.org/005b4k264"
        ▼ target-doi:  "https://doi.org/10.2807/1560-7917.ES.2021.26.5.210204e"
        ▼ reasons:
          0:           "Error in Data"
          1:           "Error in Text"
          2:           "Upgrade/Update of Prior Notice"
      update-nature:   "Correction"
      notes:           ""
```

Fig. 2 Example of contents in the "update-nature" field in Crossref metadata[10]

---

[10] https://api.labs.crossref.org/works/10.2807/1560-7917.ES.2020.25.3.2000045?mailto=sergiy.nazarovets@gmail.com

```
▼ biblio:
    volume:       "25"
    issue:        "3"
    first_page:   null
    last_page:    null
  is_retracted:   true
  is_paratext:    false
▼ primary_topic:
    id:           "https://openalex.org/T11754"
    display_name: "Diagnostic Methods for COVID-19 Detection"
    score:        0.9999
```

Fig. 3 Example of incorrect contents of the "is_retracted" field in OpenAlex metadata[11]

In a blog post, Herb (2024) highlights the issue of inaccurate representation of retractions in OpenAlex, resulting in the misclassification of papers within institutional repositories. Consequently, the ramifications of this problem extend beyond users directly accessing OpenAlex via the web interface to encompass users of other academic resources leveraging the OpenAlex API.

Given the far-reaching implications of this issue, it was imperative that it is promptly addressed. As it is of utmost importance to ensure the accurate portrayal of publication statuses on retractions, we have contacted the OurResearch team on March 19, 2023 and brought the issue to their attention. Approximately, 2300 incorrect records were identified and corrected. Metadata provided via the API between December 22, 2023 and March 19, 2024 as well as the data snapshot releases 2024-01-24 and 2024-02-27 are affected.[12]

In general, it is recommended to subject such critical metadata to a close examination, including with alternative tools for verifying the status of publications, such as the Problematic Paper Screener's Annulled Detector[13]. By adopting a multifaceted approach, stakeholders can mitigate the potential consequences of mislabelled retractions while awaiting a resolution from the OpenAlex team. Furthermore, it should be noted that every indication of a retraction status must be subject to special care. In particular, complexity reduction in the metadata representation must not lead to a loss of information as described here.

**Data availability statement**
The data and script used to generate the figures in this paper can be found at:
https://github.com/hauschke/openalex_retractions/

**Conflicts of interest**
The authors declare no relevant conflicts of interest.

**Funding statement**
No funding was received by the authors or for this research. The publication of this article was funded by the Open Access Fund of Technische Informationsbibliothek (TIB).

---

[11] https://api.openalex.org/works/W3001195213
[12] https://groups.google.com/g/openalex-users/c/y8FeQR9UhAQ
[13] https://www.irit.fr/~Guillaume.Cabanac/problematic-paper-screener

**References**

COPE Council. (2019). COPE retraction guidelines. https://doi.org/10.24318/cope.2019.1.4

Corman, V. M., Landt, O., Kaiser, M., Molenkamp, R., Meijer, A., Chu, D. K., et al. (2020). Detection of 2019 novel coronavirus (2019-nCoV) by real-time RT-PCR. *Eurosurveillance*, *25*(3). https://doi.org/10.2807/1560-7917.ES.2020.25.3.2000045

Else, H., & Van Noorden, R. (2021). The fight against fake-paper factories that churn out sham science. *Nature*, *591*(7851), 516–519. https://doi.org/10.1038/d41586-021-00733-5

Freiermuth, M. R. (2023). Now you have to pay! A deeper look at publishing practices of predatory journals. *Learned Publishing*, *36*(4), 667–688. https://doi.org/10.1002/leap.1583

Harder, R. (2024). Using Scopus and OpenAlex APIs to retrieve bibliographic data for evidence synthesis. A procedure based on Bash and SQL. *MethodsX*, *12*, 102601. https://doi.org/10.1016/j.mex.2024.102601

Herb, U. (2024). Retractions in Web of Science and OpenAlex. *ScholComm, Ulrich's Notes*. https://scidebug.com/2024/01/31/retractions-in-web-of-science-and-openalex/

Hesselmann, F., Graf, V., Schmidt, M., & Reinhart, M. (2017). The visibility of scientific misconduct: A review of the literature on retracted journal articles. *Current Sociology*, 65(6), 814–845. https://doi.org/10.1177/0011392116663807

Joelving, F. (2024). Paper trail. *Science*, *383*(6680), 252–255. https://doi.org/10.1126/science.zrjehzt

Jahn, N., Haupka, N. & Hobert, A. (2023). Analysing and reclassifying open access information in OpenAlex. *Scholarly Communication Analytics*. https://subugoe.github.io/scholcomm_analytics/posts/oalex_oa_status/

Kendall, G., & Teixeira da Silva, J. A. (2024). Risks of abuse of large language models, like ChatGPT, in scientific publishing: Authorship, predatory publishing, and paper mills. *Learned Publishing*, *37*(1), 55–62. https://doi.org/10.1002/leap.1578

Liu, L., & Liu, W. (2023). The engagement of academic libraries in open science: A systematic review. *The Journal of Academic Librarianship*, *49*(3), 102711. https://doi.org/10.1016/j.acalib.2023.102711

Mallapaty, S. (2024). China conducts first nationwide review of retractions and research misconduct. *Nature*, *626*(8000), 700–701. https://doi.org/10.1038/d41586-024-00397-x

Piniewski, M., Jarić, I., Koutsoyiannis, D., & Kundzewicz, Z. W. (2024). Emerging plagiarism in peer-review evaluation reports: a tip of the iceberg? *Scientometrics*, (0123456789). https://doi.org/10.1007/s11192-024-04960-1

Priem, J., Piwowar, H., & Orr, R. (2022). OpenAlex: A fully-open index of scholarly works, authors, venues, institutions, and concepts. https://arxiv.org/abs/2205.01833.

Rivera, H., & Teixeira da Silva, J. A. (2021). Retractions, Fake Peer Reviews, and Paper Mills. *Journal of Korean Medical Science*, *36*(24), 1–7. https://doi.org/10.3346/jkms.2021.36.e165

Scheidsteger, T., & Haunschild, R. (2023). Which of the metadata with relevance for bibliometrics are the same and which are different when switching from Microsoft Academic Graph to OpenAlex? *El Profesional de la información*. https://doi.org/10.3145/epi.2023.mar.09

Singh Chawla, D. (2022). Massive open index of scholarly papers launches. *Nature*. https://doi.org/10.1038/d41586-022-00138-y

Velez-Estevez, A., Perez, I. J., García-Sánchez, P., Moral-Munoz, J. A., & Cobo, M. J. (2023). New trends in bibliometric APIs: A comparative analysis. *Information Processing & Management*, *60*(4), 103385. https://doi.org/10.1016/j.ipm.2023.103385

Vuong, Q.-H. (2020). Reform retractions to make them more transparent. *Nature*, *582*(7811), 149–149. https://doi.org/10.1038/d41586-020-01694-x

Zhang, L., Cao, Z., Shang, Y., Sivertsen, G., & Huang, Y. (2024). Missing institutions in OpenAlex: possible reasons, implications, and solutions. *Scientometrics*. https://doi.org/10.1007/s11192-023-04923-y